%% file: fs20.tex
\documentclass[11pt,aps,pra,tightenlines,nofootinbib,notitlepage,longbibliography]{revtex4-1}
\pdfoutput=1

\usepackage{amsmath}
\usepackage{amssymb}
\usepackage{bbm}
\usepackage{simplewick}
\usepackage{mathtools}
\usepackage[normalem]{ulem}
\usepackage{dsfont}
\usepackage{mathrsfs}
\usepackage[makeroom]{cancel}
\usepackage{graphicx}

\input{sheader}

\usepackage[ddmmyyyy]{datetime}
\makeatletter
\def\Dated@name{}
\makeatother

\begin{document}

\title{Electromagnetic angular momentum of the electron: One-loop studies
}
\author{Bogdan Damski}
\affiliation{Jagiellonian University, Marian Smoluchowski Institute of Physics, {\L}ojasiewicza 11, 30-348 Krak\'ow, Poland}
\begin{abstract}
We study angular momentum of the electron stored in its electric and magnetic fields. 
We use for this purpose  quantum electrodynamics in the covariant gauge. 
We show that a finite
 one-loop result for such angular momentum can be  obtained without invoking any renormalization procedure.
 We compare it to the  classical estimation relying on a short-distance cutoff.
\end{abstract}
\maketitle

\section{Introduction}
\label{Introduction_sec}

When electric ($\E$) and magnetic ($\Bold$) fields cross, the Poynting
vector, $\E\times\Bold$, 
tells us that there is a flow of electromagnetic energy.
Angular momentum associated with it reads \cite{JacksonBook}
\be
\J_\fl=\int\d{r}\r\times(\E\times\Bold)
\label{Jclass}
\ee
and we will call it field angular momentum for brevity.

Such a form of  angular momentum is quite intriguing   if we
notice that it is generically non-zero  in static electromagnetic fields,
where no dynamics seems to be happening at first glance. For example,
a charge and a magnet  placed at fixed-in-time positions create
all around a ``circular'' flow
of the electromagnetic energy  producing non-zero angular momentum  density.
As $\J_\fl$ is  a  part of  total angular momentum, its changes in systems,
where total angular momentum  is conserved, induce changes in  angular momentum
associated with other degrees of freedom (e.g. a much more intuitive
mechanical angular momentum). 
A famous   example of this phenomenon
is known as  the  Feynman's disk paradox, 
where one considers 
 an electrically charged disk and a superconducting
wire carrying an  electric current (see Secs. 17-4 and 27-6 of \cite{FeynmanVolII} and 
\cite{LombardiAmJPhys1983,BahderAmJPhys1985,MaAmJPhys1986}).
When  temperature rises, the 
current disappears and the disc starts rotating. This   seemingly violates  angular momentum
conservation if one forgets about conversion  of  vanishing   field angular momentum
into  mechanical angular  momentum of the disc.
It is thus reasonable  to argue that  $\J_\fl$ is a   fundamentally-important
counterintuitive quantity deserving  in-depth theoretical and experimental studies. 

One of the simplest  settings  for its discussion  is found by considering  a  physical  object 
at rest   having the charge $q$ and the magnetic moment $\bmu$. Far away from it, where 
not only details of its structure but also quantum effects can be neglected, its electric and magnetic fields are
well-approximated by classical expressions \cite{JacksonBook}
\be
\E=\frac{q\r}{4\pi r^3}, \ \ \Bold=\frac{3(\bmu\cdot\hat\r)\hat\r-\bmu}{4\pi r^3},
\label{EBclass}
\ee
where $r=|\r|$ and $\hat\r=\r/r$.  The position vector $\r$ goes from the object to the point, where the
fields are discussed.

\begin{figure}[t]
\includegraphics[width=0.3\textwidth,clip=true]{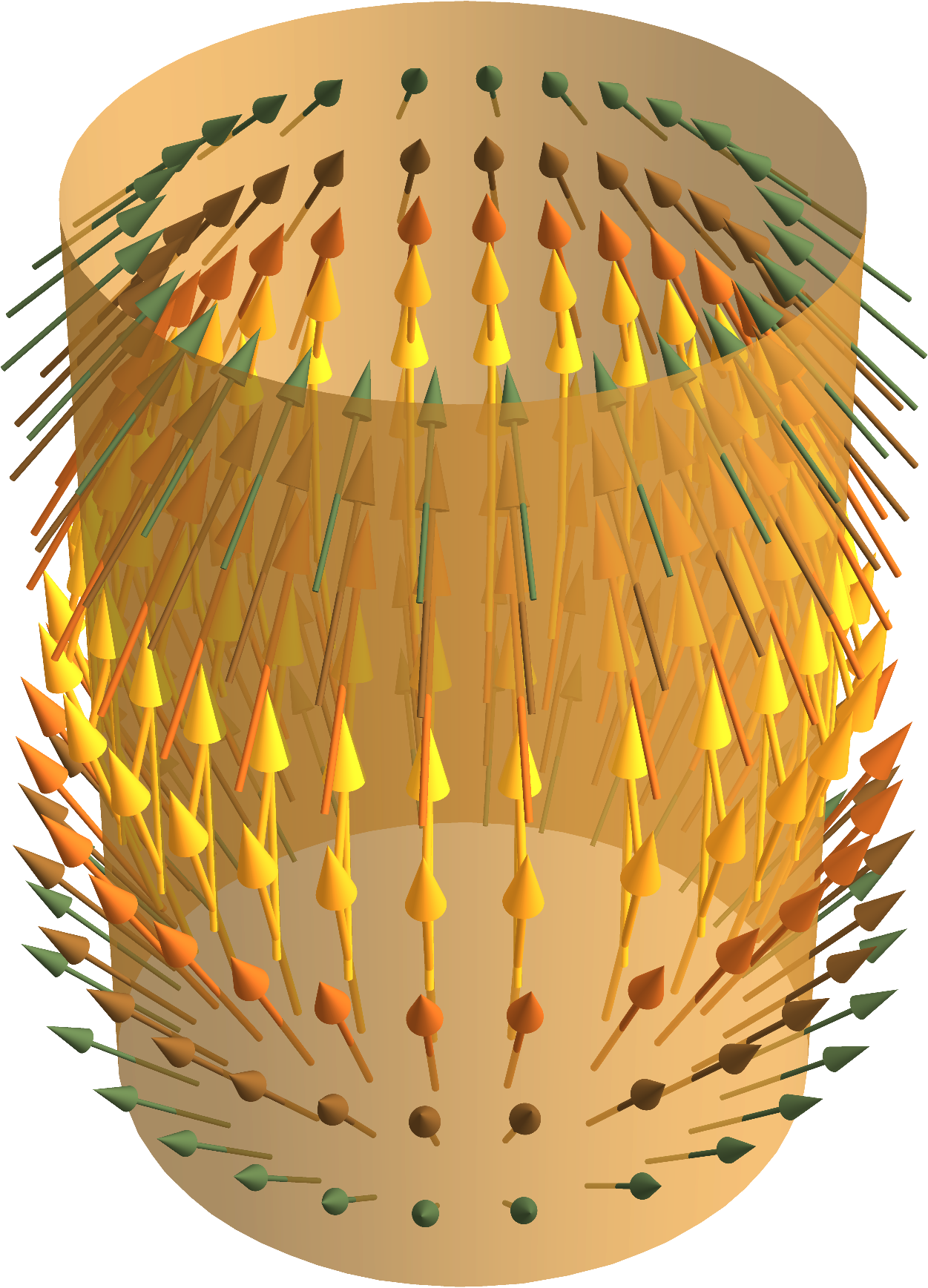}
\caption{Density of angular
momentum of  electromagnetic fields (\ref{EBclass}). 
For the clarity of presentation, we show $r^4\JJ_\fl$ 
on the surface  $x^2+y^2=\text{const}$, where $\r=(x,y,z)$,
the $z=0$ plane cuts the cylinder in half, and $\JJ_\fl=\r\times(\E\times\Bold)$. 
The magnetic moment $\bmu$ is
anti-aligned with the $z$-axis,  it points downward  the cylinder. The charge $q<0$.
}
\label{zeb_fig}
\end{figure}

A quick look at density of such angular momentum, which we depict in Fig.
\ref{zeb_fig}, shows that one can anticipate a non-trivial result for field angular momentum.
To quantify this expectation, one  restricts the integration in (\ref{Jclass}) to $r\ge r_c$, where $r_c$
is large-enough to ensure that the use of    (\ref{EBclass})  is justified.
It is then a simple exercise to show that \cite{HigbieAmJPhys1988}
\be
\J_\fl = \bmu \frac{q}{6\pi r_c}.
\label{Jclass1}
\ee
Two remarks  are in order now.

First, $\J_\fl$  is parallel (anti-parallel)  to the magnetic moment for positively
(negatively) charged objects. The same relation between  spin angular
momentum and the  magnetic moment is observed for  protons and electrons.

Second, it is of interest to find  what result could be obtained if one employs  some classically-motivated
value for the cutoff $r_c$ \cite{HigbieAmJPhys1988}. 
A characteristic length-scale  that can be used for such a 
purpose exists within the century-old classical
theory of the electron (see e.g. \cite{RohrlichBook1973}). 
It is known as the classical electron radius  
\be
r_0=\frac{e^2}{4\pi m},
\ee
where $m$ is the electron's mass and  $e<0$ is its charge. Leaving aside for
the moment the
question of whether it is justified to use (\ref{EBclass}) at such a short
distance from the charge, one may  take the ``classical'' electron as our physical  object, set
$q=e$,  and assume that $r_c=O(r_0)$. All this results in  
\be
\J_\fl=O\B{\bmu \frac{m}{e}}
\label{Jclass2}
\ee
suggesting that  field angular momentum of the ``classical'' electron is of
the order of  electron's spin  if one additionally takes into account that $|\bmu|$  
is of the order of the Bohr magneton 
\be
\frac{|e|}{2m}
\ee
for the  electron.
While such an estimation  clearly cannot be treated too seriously, it is
 interesting to  set it against the outcome of a  fully quantum calculation.

The purpose of this work is to compute  field angular momentum of the
electron in the framework of quantum electrodynamics (QED).
Such  a calculation not only comprehensively accounts for the quantum effects, but it also does not
rely on a  short-distance cutoff. It is therefore  interesting to inquire, and in fact
a priori unknown, whether the result of such a calculation will be finite.
We find it thus remarkable that  a finite non-trivial result
for such a physically interesting quantity can be obtained. It  comes from   our  one-loop
calculation, which  does not involve any renormalization procedure. 
There are different ways how one  can place this result in a wider context.

On the one hand,  it  provides one more  
physical  quantity characterizing the electron, arguably one of the
most important subatomic  particles. In some sense, such a result
is similar  to the Schwinger's prediction for the electron's anomalous magnetic moment, 
which also  comes from a  one-loop  calculation  and provides a basic
insight into the properties of the electron.

On the other hand, our work can be seen as a part of a larger  program
targeting characterization of  all components of  angular
momentum of the electron.
So far there have been only a few attempts in this direction \cite{BurkardtPRD2009,LiuPRD2015,JiPRD2016,BDspin},
and none of them   studied field angular momentum that we discuss here.
A similar program is being carried out for nucleons, where various
calculations have been set against  experimental data (see e.g. 
\cite{LeaderPhysRep2014,Deur2019} for  recent review articles).

The outline of this paper is the following. 
We briefly explain in Sec. \ref{Basic_sec} how our calculations will be carried
out. The actual computations are presented in 
Sec. \ref{Field_sec}, where we study   field angular momentum of the electron 
with the help of three-dimensional (3D) cutoff  and  Pauli-Villars regularizations. 
Our results are then discussed in Sec. \ref{Discussion_sec}.
The paper ends with Appendix \ref{Conventions_sec}, where 
our conventions  are briefly summarized.

\section{Basic equations}
\label{Basic_sec}
We start with the QED   Lagrangian density 
\be
\begin{aligned}
{\cal L} =& -\frac{1}{4} F_{\mu\nu}F^{\mu\nu}
+ \overline{\psi}\B{\ii\gamma^\mu\partial_\mu-
\mo}\psi- e_\o\overline{\psi}\gamma^\mu\psi A_\mu,
\end{aligned}
\label{LL}
\ee
where $\mo$ and $\eo<0$ are the bare mass and charge of the electron,
respectively,  and the
electromagnetic and fermionic fields are defined as always \cite{Greiner}. 
The theory is  canonically quantized in the standard way  after adding the
gauge-fixing term 
\be
-\frac{1}{2}\B{\partial_\mu A^\mu}^2
\ee
to the Lagrangian density \cite{Greiner}. Such a choice leads to 
Feynman-gauge QED (we will argue below that our results are the same in any
covariant gauge). 
We also mention that all  fields from now on
will be either Heisenberg-  or interaction-picture operators. 
The latter will be distinguished from the former by the index $I$.

Next, we replace the classical fields in (\ref{Jclass}) with operators,   impose normal ordering, which we denote by 
$\N{ \ }$, and rewrite the resulting expression to get
\be
J^i_\fl=\int\d{z} \varepsilon^{imn} z^m \N{F_{0j}F_{jn}}.
\label{Jfl}
\ee
We call (\ref{Jfl})  the  field angular momentum operator. 
It appears in the so-called Belinfante \cite{BelinfantePhys1939} 
and Ji \cite{JiPRL1997} decompositions of  total angular momentum,
where it has  the physical interpretation of the photon total angular
momentum operator \cite{LeaderPhysRep2014}--see this reference also for a comprehensive discussion  of
different total angular momentum decompositions in QED.   
This operator is gauge
invariant and so its expectation value should be  measurable in principle.
Such an important  property should not be taken for granted. Indeed, 
in the so-called Jaffe-Manohar \cite{JaffeNPB1990} decomposition of  total angular momentum,
the following operator accounts for  angular momentum of the
electromagnetic field
\be
\int \d{z}\varepsilon^{imn}\N{F_{m0}A_n} +  \int\d{z} \varepsilon^{imn}z^m \N{F_{j0}\partial_n A_j},
\label{Jkan}
\ee
where the first (second) term has the physical interpretation of the photon spin
(orbital) angular momentum operator \cite{LeaderPhysRep2014}. Operator (\ref{Jkan}) is gauge non-invariant in the
presence of charges, which can be easily checked. 
Differences between (\ref{Jfl}) and (\ref{Jkan})  nicely illustrate the fact that in the interacting theory such
as QED, where electromagnetic and fermionic fields are coupled,
there is no unique division of  total angular momentum into 
electromagnetic and fermionic components. Still, the study of (\ref{Jfl}) 
is  well-motivated physically and it provides a finite  gauge invariant result  relevant for 
understanding of angular momentum of the electron within the Belinfante
and Ji decompositions.

We will compute expectation value of (\ref{Jfl}) 
with the help of the  bare perturbation theory in the QED ground state $\Oprim$
with one  net electron (the difference between the number of electrons and
positrons in such a state is $+1$).
For this  purpose, we will use imaginary time evolution starting
from the ground state of the non-interacting theory with one electron at rest
 having  spin projection onto the $z$-axis  
\be
s_z=\pm\frac{1}{2}.
\label{sz}
\ee
We refer to  such a state as $|\0 s\ra$ and mention that the expectation value of
the total angular momentum operator in states $|\0s\ra$ and  $\Oprim$ equals $\spinz$.

Adopting the results of \cite{PS} to our problem, we get 
\be
\exval{\J_\fl}{\Opr}=\limT\frac{\la\0s|\T\J^I_\fl\exp(-\ii\intT\dd{x}{\cal H}^I_\IN)|\0s\ra}{
\la\0s|\T\exp(-\ii\intT\dd{x}{\cal H}^I_\IN)|\0s\ra}, 
\label{fg}
\ee
where 
\begin{gather}
\exval{\cdots}{\Opr}=\frac{\la\Opr|\cdots|\Opr\ra}{\la\Opr|\Opr\ra},\\
{\cal H}^{I}_\IN(x)= \eo \N{\overline{\psi}_I(x)\gamma^\mu\psi_I(x)}A^{I}_\mu(x),\\
\intT\dd{x}=\intTT dx^0\int\d{x},
\end{gather}
and $\T$ is the time-ordering operator.

Equation (\ref{fg}) can be substantially simplified. Indeed, with the help of our results
presented in \cite{BDspin},
it can be  rigorously shown that one can safely do the replacement
\be
\limT\int_T\dd{x}\to\int\dd{x}
\label{qwwww}
\ee
 if the  calculations of field angular momentum of the electron are infrared-regularized.
This  leads to 
\be
\exval{\J_\fl}{\Opr}=
\frac{\la\0s|\T\J^I_\fl\exp(-\ii\int\dd{x}{\cal H}^I_\IN)|\0s\ra}{
\la\0s|\T\exp(-\ii\int\dd{x}{\cal H}^I_\IN)|\0s\ra}
\label{fghj}
\ee
in accordance with the standard textbook description of the imaginary time
evolution technique \cite{PS}. Replacement (\ref{qwwww}), however,  should not be taken for
granted, which we illustrate in \cite{BDspin}.

Finally, we need the interaction-picture version of $\J_\fl$, which is
obtained by replacing the Heisenberg-picture  operators with their
interaction-picture counterparts. This can be seen by using   canonical commutation relations 
\be
[A_\mu(x^0,\y),A_\nu(x)]=0
\label{canonical}
\ee
to show that the last term of 
\be
\B{\partial_\mu A_\nu(x)}_I=\partial_\mu A_\nu^I(x) + 
\ii\eta_{\mu0}\int\d{y}\B{[{\cal H}_\IN(x^0,\y),A_\nu(x)]}_I
\ee
vanishes  (see e.g. \cite{Greiner} for the transformation relating the two
pictures).

\section{Field angular momentum}
\label{Field_sec}
We will compute here the expectation value of the field angular momentum
operator using two regularization methods. Most of the computations in this section, however, can be
done without referring to any regularization technique. Such results will be 
collected in Sec. \ref{Base_sec}. They will be then adapted to calculations 
based on either the 3D
cutoff (Sec. \ref{3D_sec}) or Pauli-Villars (Sec. \ref{Pauli_sec})
regularization.

\subsection{Base formulae}
\label{Base_sec}
To calculate the expectation value of field angular momentum operator
(\ref{Jfl}), we expand (\ref{fghj}) in the series in $\eo$   
\be
\exval{J^i_\fl}{\Opr}=-\frac{1}{2\Vol}\int\dd{x}\dd{y}
\la\0s|\T(\J^I_\fl)^i {\cal H}^{I}_\IN(x){\cal H}^{I}_\IN(y)|\0s\ra+O(\eo^4),
\label{Jpert_fl}
\ee
where we have replaced the denominator of (\ref{fghj}) with 
\be
\Vol=\la\0s|\0s\ra=\int \frac{\d{x}}{(2\pi)^3}
\label{VV}
\ee
because we work in the quadratic order in $\eo$ and $\la\0s|\J^I_\fl|\0s\ra=0$. 
Normalization factor (\ref{VV}) of  delocalized  one-electron states  is 
formally infinite, but it unambiguously cancels down during calculations
(see e.g. discussion in \cite{BDspin}).  
This is a well-known feature of  calculations done in the plane-wave basis.

The electromagnetic and fermionic contributions, to the  matrix element in (\ref{Jpert_fl}), factor out
\begin{subequations}
\begin{align}
\label{EF}
&\la\0s|\T(\J^I_\fl)^i {\cal H}^{I}_\IN(x){\cal H}^{I}_\IN(y)|\0s\ra=
\eo^2{\cal E}^i_{\mu\nu}(x,y){\cal F}^{\mu\nu}(x,y),\\
\label{Aimunu}
&{\cal E}^i_{\mu\nu}(x,y)=\varepsilon^{imn}\int\d{z}z^m\langle0|\T\N{F^I_{0j}(z) F^I_{jn}(z)}  A^I_\mu(x)A^I_\nu(y)|0\rangle,\\
&{\cal F}^{\mu\nu}(x,y)=\la\0s|\T\N{\overline{\psi}_I(x)\gamma^\mu\psi_I(x)}\N{\overline{\psi}_I(y)\gamma^\nu\psi_I(y)}|\0s\ra,
\label{Fmunu}
\end{align}
\label{matel}%
\end{subequations}
where $|0\ra$ is the vacuum state of the non-interacting theory and 
$z^0$ is dropped  from the list of arguments of ${\cal E}^i_{\mu\nu}$ for the sake of brevity
(the same is done for ${\cal A}^i_{\mu\nu}$ and ${\cal B}^i_{\mu\nu}$ that will
be introduced below).

To compute the fermionic matrix element, we apply  the Wick's theorem to (\ref{Fmunu})
\be
\begin{aligned}
{\cal F}^{\mu\nu}(x,y)&=
\contraction{\la}{\0s}{|}{\overline{\psi}}
\contraction{\la\0s|\overline{\psi}_I(x)\gamma^\mu}{\psi}{_I(x)}{\overline{\psi}}
\contraction{\la\0s|\overline{\psi}_I(x)\gamma^\mu\psi_I(x)\overline{\psi}_I(y)\gamma^\nu}{\psi}{_I(y)|}{\0s}
\la\0s|\overline{\psi}_I(x)\gamma^\mu\psi_I(x)\overline{\psi}_I(y)\gamma^\nu\psi_I(y)|\0s\ra
+
\contraction{\frac{1}{2}\la\0s|}{\overline{\psi}}{_I(x)\gamma^\mu\psi_I(x)\overline{\psi}_I(y)\gamma^\nu}{\psi}
\contraction{\frac{1}{2}\la\0s|\overline{\psi}_I(x)\gamma^\mu}{\psi}{_I(x)}{\overline{\psi}}
\frac{1}{2}\la\0s|\overline{\psi}_I(x)\gamma^\mu\psi_I(x)\overline{\psi}_I(y)\gamma^\nu\psi_I(y)|\0s\ra\\
&+(\barexymunu \ \text{on all terms}).
\end{aligned}
\ee
This can be evaluated with the following contractions      
\begin{gather}
\label{prop_fer}
\contraction{}{\psi}{_{I}(x)}{\psi}\psi_{I}(x)\overline{\psi}_{I}(y)=
\la0|\T \psi_I(x)\overline{\psi}_I(y)|0\ra=
\ii\int\frac{\dd{p}}{(2\pi)^4}\frac{\gamma\cdot p+\mo}{p^2-\mo^2+\izero}e^{-\ii p\cdot(x-y)},\\
\contraction{\la}{\0 s}{|}{\overline{\psi}}
\la\0 s|\overline{\psi}_I(x)=\frac{\ous}{(2\pi)^{3/2}}e^{\ii f\cdot x},   \ 
\frac{\us}{(2\pi)^{3/2}}e^{-\ii f\cdot x}=
\contraction{}{\psi}{_I(x)|}{\0 s}
\psi_I(x)|\0 s\ra,
\label{ext_contractions}
\end{gather}
where 
\be
f=(\mo,\0)
\ee
and  the $\us$ bispinors, 
describing an electron at rest with the spin projection onto the $z$-axis given by (\ref{sz}), are provided in  Appendix 
\ref{Conventions_sec}.

After a few elementary steps, we arrive at
\be
\begin{aligned}
{\cal F}^{\mu\nu}(x,y)&=\frac{\ii}{(2\pi)^3}\int\dddd{p}
\frac{\ous\gamma^\mu (\gamma\cdot p+\mo)\gamma^\nu\us}{p^2-\mo^2+\izero}e^{\ii (f-p)\cdot (x-y)}\\
&+\frac{V}{2}\int \dddd{p}\dddd{q} 
\frac{\trr{(\gamma\cdot p+\mo)\gamma^\mu(\gamma\cdot q+\mo)\gamma^\nu }}{(p^2-\mo^2+\izero)(q^2-\mo^2+\izero)}
e^{\ii(p-q)\cdot(x-y)}\\
&+(\barexymunu \ \text{on all terms})
\end{aligned}
\label{FFmunu}
\ee
and, to avoid any confusion, we mention that throughout this work there is no summation over $s$ in 
bispinor matrix elements $\ous\cdots\us$
(we do not average over spin polarizations).

To simplify (\ref{FFmunu}), we need the following well-known
representation-independent identity
\be
\label{aass}
\gamma^\mu\gamma^\nu\gamma^\rho=\eta^{\mu\nu}\gamma^\rho+\eta^{\nu\rho}\gamma^\mu
-\eta^{\mu\rho}\gamma^\nu-\ii\varepsilon^{\sigma\mu\nu\rho}\gamma_\sigma\gamma^5,
\ee
where $\gamma^5=\ii\gamma^0\gamma^1\gamma^2\gamma^3$. We also need 
\begin{align}
&\trr{(\gamma\cdot p+\mo)\gamma^\mu(\gamma\cdot q+\mo)\gamma^\nu}=4\BB{p^\mu q^\nu+p^\nu
q^\mu - \eta^{\mu\nu} (p\cdot q-\mo^2)},\\
&\ous\gamma^\mu \us=\eta^{\mu0}, \ \
\ous\gamma^\mu\gamma^\nu\us=\eta^{\mu\nu}-2\ii\sz\varepsilon^{0\mu\nu3}, \ \ 
\varepsilon^{\sigma\mu\nu\rho}\ous\gamma_\sigma\gamma^5\us=2\sz\varepsilon^{\mu\nu\rho3},
\label{uueps}
\end{align}
which can be easily verified in the standard representation of $\gamma$
matrices (the same results are obtained in all 
representations, which are   unitarily  similar to the standard one: Weil, Majorana, etc.).
Having these expressions at hand, we would like to remark that field angular
momentum  will be $\sz$-dependent\footnote{The same is observed in 
(\ref{Jclass2}),  if we note that  the electron's magnetic moment also depends on  the spin projection.}.
As a result, we learn from  (\ref{uueps})  that our calculations will
critically depend on the four-dimensional Levi-Civita symbol, 
which  is troublesome in dimensional regularization (see e.g.  Appendix B.2 of \cite{DreinerPhysRep2010}). 
In fact, the 3D version of this symbol has already appeared in the field angular
momentum operator, whose definition is heavily rooted in dimensionality of
the physical space. These complications  discourage us from using dimensional
regularization in the subsequent sections. 

Combining (\ref{FFmunu})  with (\ref{aass})--(\ref{uueps}), 
the fermionic matrix element can be not only simplified but also 
 decomposed  into 
symmetric and anti-symmetric parts with
respect to the 
$\mu\leftrightarrow\nu$
transformation 
\begin{subequations}
\be
{\cal F}^{\mu\nu}(x,y)={\cal F}_\sym^{\mu\nu}(x,y)+{\cal F}_\asym^{\mu\nu}(x,y),
\label{Fsymasym}
\ee
\ba
{\cal F}_\sym^{\mu\nu}(x,y)&= \frac{\ii}{(2\pi)^3}\int \dddd{p} 
\frac{p^\mu\eta^{\nu0}+p^\nu\eta^{\mu0}-p^0\eta^{\mu\nu}
+ \mo\eta^{\mu\nu}}{p^2-\mo^2+\izero} 
e^{\ii(f-p)\cdot(x-y)}
\\
&+2V \int  \dddd{p}   \dddd{q}
\frac{p^\mu q^\nu+p^\nu q^\mu-\eta^{\mu\nu}(p\cdot q-\mo^2)}{\B{p^2-\mo^2+\izero}
\B{q^2-\mo^2+\izero}} 
e^{\ii(p-q)\cdot(x-y)}
\\
&+ (x\leftrightarrow y  \ \text{on all terms}),
\label{bbbnnnmmm}
\ea
\ba
&{\cal F}^{\mu\nu}_\asym(x,y)= \frac{2\sz}{(2\pi)^3}\int \dddd{p}
\frac{
\varepsilon^{0\mu\nu3}\mo-\varepsilon^{\sigma\mu\nu3}p_\sigma
}{p^2-\mo^2+\izero} 
e^{\ii(f-p)\cdot (x-y)}
- (x\leftrightarrow y).
\ea
\label{ploik}%
\end{subequations}

To compute the electromagnetic matrix element,  we again make use of  the Wick's theorem 
\be
\begin{aligned}
{\cal E}^i_{\mu\nu}(x,y)&=
\varimn\int\d{z}z^m
\contraction{}{F}{^I_{0j}(z) }{A}F^I_{0j}(z) A^I_\mu(x)
\contraction{}{F}{^I_{jn}(z) }{A}F^I_{jn}(z) A^I_\nu(y)
+(\barexymunu).
\end{aligned}
\label{mat_fl}
\ee
Then, we need  the  interaction-picture photon propagator in the Feynman gauge 
\be
\contraction{}{A}{^{I}_\mu(x)}{A}A^{I}_\mu(x)A^{I}_\nu(y)=
\la0|\T A^I_\mu(x)A^I_\nu(y)|0\ra
=
-\ii\int\dddd{p}\frac{\eta_{\mu\nu}}{p^2+\izero}e^{-\ii p\cdot(x-y)}
\label{prop_el}
\ee
and the identity  
\be
\la0|\T \partial_\alpha A^I_\beta(x)A^I_\gamma(y)|0\ra=
\frac{\partial}{\partial x^\alpha}\la0|\T A^I_\beta(x)A^I_\gamma(y)|0\ra,
\label{bgt}
\ee
which can be trivially proved  with   (\ref{canonical}).

Combining these simple results, we get 
\be
\contraction{}{F}{^I_{\alpha\beta}(z) }{A}F^I_{\alpha\beta}(z)
A^I_\gamma(x)=\int\dddd{p}
\frac{p_\alpha\eta_{\beta\gamma}-p_\beta\eta_{\alpha\gamma}}{p^2+\izero}
e^{-\ii p\cdot (x-z)}.
\label{FAcont}
\ee
Quite interestingly, if we would use the general covariant gauge photon
propagator \cite{Lautrup1967,Greiner}, which is obtained by replacement 
\be
\eta_{\mu\nu}\to \eta_{\mu\nu}+\xxii\frac{p_\mu p_\nu}{p^2+\izero}
\ee
in (\ref{prop_el}), we would get the same result for (\ref{FAcont}) for all
parameters $\xi$ labeling various covariant gauge choices.
This shows that  our results are gauge independent within the family of all
covariant gauges, which is a welcome feature.

Using (\ref{FAcont}) to evaluate (\ref{mat_fl}), we obtain
\begin{subequations}
\begin{align}
&{\cal E}^i_{\mu\nu}(x,y)=\varimn\int\d{z}z^m
\int\dddd{p}\dddd{q} 
\frac{\Delta_{\mu\nu,n}(p,q)+\Delta_{\nu\mu,n}(q,p)}{(p^2+\izero)(q^2+\izero)} 
e^{-\ii p\cdot x +\ii q\cdot y+\ii (p-q)\cdot z},\\
&\Delta_{\mu\nu,n}(p,q)=(p_0\eta_{j\mu}-p_j\eta_{0\mu})(q_n\eta_{j\nu}-q_j\eta_{n\nu}).
\end{align}
\label{eeee}%
\end{subequations}

Next, we use 
\be
\int\ddd{z}z^m e^{\ii \z\cdot(\q-\p)} = \frac{\ii}{2}\B{\frac{\partial}{\partial p^m} - \frac{\partial}{\partial
q^m}}\delta(\p-\q),
\label{ddelta}
\ee
and integrate  by parts. As can be easily checked, such integration by parts  does not generate
boundary contributions.

Finally, introducing 
\be
\tilde q=(q^0,\p),
\ee
we  derive 
\begin{subequations}
\be
{\cal E}^i_{\mu\nu}(x,y)={\cal A}^i_{\mu\nu}(x,y) + {\cal B}^i_{\mu\nu}(x,y),
\ee
\be
{\cal A}^i_{\mu\nu}(x,y)=\frac{1}{2}\varepsilon^{imn}
(x^m+y^m) 
\int \dddd{p}\frac{dq^0}{2\pi} 
\frac{\Delta_{\mu\nu,n}(p,\tilde q)+\Delta_{\nu\mu,n}(\tilde q,p)}{(p^2+\izero)(\tilde q^2+\izero)}
e^{-\ii p\cdot x+\ii \tilde q\cdot y  + \ii  (p^0-q^0)z^0},
\ee
\begin{multline}
{\cal B}^i_{\mu\nu}(x,y)=
-\frac{\ii}{2}
\varepsilon^{imn}
\int \dddd{p}\frac{dq^0}{2\pi} 
e^{-\ii p\cdot x+\ii \tilde q\cdot y + \ii  (p^0-q^0) z^0}\\
\cdot\left.
\B{\frac{\partial}{\partial p^m}-\frac{\partial}{\partial q^m}}
\frac{\Delta_{\mu\nu,n}(p,q)+\Delta_{\nu\mu,n}(q,p)}{(p^2+\izero)(q^2+\izero)}\right|_{\q=\p}. 
\end{multline}
\label{Eall}%
\end{subequations}

We get by collecting  (\ref{Jpert_fl}), (\ref{matel}), (\ref{ploik}), and (\ref{Eall})
\be
\exval{J^i_\fl}{\Opr}=-\frac{\eo^2}{2V}\int\dd{x}\dd{y}
\BB{{\cal A}^i_{\mu\nu}(x,y)+{\cal B}^i_{\mu\nu}(x,y)}{\cal
F}^{\mu\nu}(x,y)+O(\eo^4),
\ee
where  the ${\cal A}^i_{\mu\nu}$ term can be dropped
because it leads to the integral of the form 
\be
\int \d{x}\d{y}(\x+\y)^me^{\ii \P\cdot(\x-\y)}=0
\ee
with  $\P$ being some combination of $3$-momenta.

In the end, we arrive at the unregularized expression for field angular momentum
of the electron
\be
\exval{J^i_\fl}{\Opr}=-2\ii\eo^2\sz
\int\dddd{p}\frac{\delta^{i3}[2(p^0-\mo)^2+\om{p}^2]-p_ip_3}{
(p^2-\mo^2+\izero)[(p-f)^2+\izero]^2}+O(\eo^4),
\label{kloi}
\ee
where $\om{p}=|\p|$. This expression,  unlike (\ref{Eall}),  is time, i.e., $z^0$-independent. It
is an anticipated  feature 
because that expectation value is computed in an eigenstate of the system 
and  $\J_\fl$ has no explicit time dependence. 
We also note that   field angular momentum of the electron 
does not have the $\sz$-independent component.
This can be explained from two different viewpoints.

First,  such a component can arise only from 
 the symmetric part of fermionic matrix
element (\ref{bbbnnnmmm}). During evaluation of 
$\int\dd{x}\dd{y}{\cal B}^i_{\mu\nu}(x,y){\cal F}^{\mu\nu}_\sym(x,y)$, however,
one encounters    contractions between 
symmetric and anti-symmetric in $\mu\leftrightarrow\nu$ tensors, which make
such an integral equal to zero.
Second,   after averaging over spin projections, 
field angular momentum of the electron should vanish and so its
$\sz$-independent component cannot exist. It is so  because after 
such an operation, there is no
preferred direction in the three-dimensional real space. 

Until now, we have gone quite far  without using  any regularization. To assign a value to 
expression (\ref{kloi}),  we need to specify a regularization scheme.  We will
discuss two options below.

\subsection{3D cutoff regularization}
\label{3D_sec}

The idea  here is to regularize calculations from Sec. \ref{Base_sec} by
cutting off  $3$-momenta in expressions for propagators.
This can be achieved by the following modification of electromagnetic
propagator (\ref{prop_el})
\be
\label{[]}
\dd{p}\to\DD{p}=
\dd{p}\theta(\Lambda_c-\om{p})\theta(\om{p}-\lambda_c),
\ee
where $\theta$ is the Heaviside step function and the infrared (IR) and ultraviolet (UV) cutoffs are denoted as
$\lambda_c$ and $\Lambda_c$, respectively.
Alternatively, one may implement  the UV cutoff in  fermionic propagator (\ref{prop_fer}) while keeping the IR
one in the electromagnetic propagator (application  of the IR cutoff to
the fermionic propagator is questionable as our imaginary time evolution starts
from the zero-momentum state).

If we redo the calculations from Sec. \ref{Base_sec} with either of the
above-outlined options, we will find that 
\begin{subequations}
\begin{align}
&\exval{\J_\fl}{\Opr}=\limlambdyc\expval{\J_\fl}{\Opr}{\lambda_c\Lambda_c}+O(\eo^4),\\
&\expval{J^i_\fl}{\Opr}{\lambda_c\Lambda_c}=-2\ii \eo^2\spinz
\int\frac{\DD{p}}{(2\pi)^4}\frac{2(p^0-\mo)^2+(p_1)^2+(p_2)^2}{(p^2-m_\o^2+\izero)[(p-f)^2+\izero]^2}.
\label{kloi1b}%
\end{align}
\label{kloi1}%
\end{subequations}

To evaluate (\ref{kloi1b}), we first  integrate over $p^0$ using the residue
theorem and then do the integration in the 3D $\p$-space. 
The order of angular and radial integrations in the $\p$-space does not matter since the two
operations commute when the radial integration is done on a bounded interval.
If that would not be the case, then the radial  integration, when 
performed  before the angular one, would produce 
a meaningless logarithmically divergent result. 

Leaving the radial integration for the last step of evaluation of (\ref{kloi1b}), we find 
\be
\expval{J^i_\fl}{\Opr}{\lambda_c\Lambda_c}=
-\frac{\eo^2\spinz}{6\pi^2}\int_{\lambda_c}^{\Lambda_c} d\om{p}
\frac{\mo^2}{\vareps{p}(\om{p}+\vareps{p})^2},
\label{o111}%
\ee
where  $\vareps{p}=\sqrt{\mo^2+\om{p}^2}$. 
This   can be computed after changing the integration  variable to $y$
given by  (see e.g. Sec. 2.25 of \cite{Ryzhik}) 
\be
y=\B{\om{p}/\mo+\sqrt{1+(\om{p}/\mo)^2}}^{-2},
\label{Ram}
\ee
which turns   the integral in  (\ref{o111}) into 
\be
\int_{y(\Lambda_c)}^{y(\lambda_c)}\frac{dy}{2},   
\ee
where $y(\om{p})$ is given by the right-hand side of (\ref{Ram}).
In the end, we  get 
\be
\exval{J^i_\fl}{\Opr}=-\spinz\frac{\eo^2}{12\pi^2}+O(\eo^4).
\label{qqqmmm}
\ee

\subsection{Pauli-Villars regularization}
\label{Pauli_sec}
We will employ the Pauli-Villars regularization in this section
\cite{PauliRMP1949}. In its simplest
version \cite{Schwartz,Gupta}, this is systematically done by  modifying the  Lagrangian density
 so that it reads 
\be
\begin{aligned}
{\cal L}=&-\frac{1}{4}F_{\mu\nu}F^{\mu\nu}   
    -\frac{1}{2}\B{\partial_\mu A^\mu}^2
   +\frac{\mph^2}{2}A_\mu A^\mu
   +\overline{\psi}\B{\ii\gamma^\mu\partial_\mu-\mo}\psi\\
   &+\frac{1}{4}\tilde F_{\mu\nu}\tilde F^{\mu\nu}
    +\frac{1}{2}\B{\partial_\mu\tilde A^\mu}^2
   -\frac{\Lambda^2}{2}\tilde A_\mu\tilde A^\mu
   +\overline{\tilde\psi}\B{\ii\gamma^\mu\partial_\mu-\Lambda}\tilde\psi\\
&-\eo (\overline{\psi}\gamma^\mu\psi + \overline{\tilde\psi}\gamma^\mu\tilde\psi)(A_\mu+\tilde A_\mu),
\label{PVL}
\end{aligned}
\ee
where  $\tilde\psi$ and $\tilde A_\mu$ are the Pauli-Villars bosonic ghost fields and
the mass term has been added for real photons.

The IR regularization is controlled  by 
 $\lambda\ll\mo$ entering the electromagnetic propagator, which now reads 
\be
\contraction{}{A}{^{I}_\mu(x)}{A}A^{I}_\mu(x)A^{I}_\nu(y)
=-\ii\int\dddd{p}\frac{\eta_{\mu\nu}}{p^2-\mphh+\izero}e^{-\ii p\cdot(x-y)},
\label{yhnnhy1}
\ee
 while the UV regularization is supposed to be controlled by $\Lambda\gg\mo$.

To see if the latter really happens,  we
replace ${\cal H}^I_\IN$ in (\ref{Jpert_fl}) with 
\be
\eo (\N{\overline{\psi}_I\gamma^\mu\psi_I} + \N{\overline{\tilde\psi}_I\gamma^\mu\tilde\psi_I})(A^I_\mu+\tilde A^I_\mu)
\ee
and redefine $|\0 s\ra$ so that it is the state with one real electron at rest in the spin state $s$ and 
zero real photons and ghost particles.
The resulting expression for field angular momentum of the electron depends on the product of ``electromagnetic''
\begin{multline}
\la0|
\T(\J^I_\fl)^i  A^I_\mu(x)A^I_\nu(y) +    
\T(\J^I_\fl)^i  A^I_\mu(x)\tilde A^I_\nu(y) \\+ 
\T(\J^I_\fl)^i  \tilde A^I_\mu(x)A^I_\nu(y) +   
\T(\J^I_\fl)^i  \tilde A^I_\mu(x)\tilde A^I_\nu(y)   |0\ra
\label{elelel}
\end{multline}
and ``fermionic'' 
\begin{multline}
\la\0s|
\T\N{\overline{\psi}_I(x)\gamma^\mu\psi_I(x)}\N{\overline{\psi}_I(y)\gamma^\nu\psi_I(y)}
+\T\N{\overline{\psi}_I(x)\gamma^\mu\psi_I(x)}\N{\overline{\tilde\psi}_I(y)\gamma^\nu\tilde\psi_I(y)}\\
+\T\N{\overline{\tilde\psi}_I(x)\gamma^\mu\tilde\psi_I(x)}\N{\overline{\psi}_I(y)\gamma^\nu\psi_I(y)}
+\T\N{\overline{\tilde\psi}_I(x)\gamma^\mu\tilde\psi_I(x)}\N{\overline{\tilde\psi}_I(y)\gamma^\nu\tilde\psi_I(y)}
|\0s\ra
\label{ferfer}
\end{multline}
matrix elements just as (\ref{Jpert_fl}) combined with (\ref{matel}) does.
The problem now is that it 
is {\it not} regularized in the UV sector. To see this, we take a
close look at (\ref{elelel}) and (\ref{ferfer}).

In the former  matrix element, the second and the third term vanishes because there is 
an odd number of real and ghost
fields and there are no contractions between them. The fourth one also
vanishes, because $\J^I_\fl$ is normal ordered. As a result, we are left with
the first term and so 
(\ref{elelel}) is the same as unregularized   (\ref{Aimunu}) if we neglect the difference
between (\ref{prop_el}) and (\ref{yhnnhy1}), which does not
provide the UV regularization that we look for.

In the latter matrix element, the second and  the third term vanishes because the
ghost operators are normal ordered. The fourth term does not vanish, but it is
independent of the spin orientation because there are no contractions of ghost
fields on states without ghost particles. Thus, it cannot 
 regularize the $\sz$-dependent final result for field angular momentum of the
 electron. In fact, by knowing that the ghost fermionic propagator is
 given by (\ref{prop_fer}) with $\mo$ replaced by $\Lambda$
 \cite{Schwartz,Gupta}, one can easily check   that
 contribution of the fourth term to the final result vanishes for the very same
 reason why ${\cal F}_\sym^{\mu\nu}$ does not contribute to
 (\ref{kloi}). So, after dropping this term, (\ref{ferfer}) is the same as
 unregularized (\ref{Fmunu}) if we note that (\ref{prop_fer}) still holds for 
 Lagrangian density (\ref{PVL}).

Therefore, we are left with the option of a formal modification of 
 propagators in the spirit of the  Pauli-Villars regularization. 
Such an approach  comes in different flavors. For example, one
can modify the electromagnetic propagator through 
\be
\frac{1}{p^2-\mphh+\izero}\to
\frac{1}{p^2-\mphh+\izero}-\frac{1}{p^2-\Lambda^2+\izero}.
\ee
Alternatively, one may modify the fermionic propagator  through either 
\be
\frac{\gamma\cdot p+\mo}{p^2-\mo^2+\izero}\to(\gamma\cdot p+\mo)
\B{
\frac{1}{p^2-\mo^2+\izero}- \frac{1}{p^2-\Lambda^2+\izero}
}
\label{regeasy}
\ee
 or   
\be
\frac{\gamma\cdot p+\mo}{p^2-\mo^2+\izero}\to
\frac{\gamma\cdot p+\mo}{p^2-\mo^2+\izero} -
\frac{\gamma\cdot p+\Lambda}{p^2-\Lambda^2+\izero},
\ee
where, e.g.,  the former option is discussed in \cite{BogolubovBook} while the latter 
one in \cite{ColemanBook}.
We have checked that those  three ways of regularization 
lead to the same final result. Therefore, we will 
employ (\ref{regeasy}) as it yields  the simplest analytical
expressions. The Pauli-Villars-regularized one-loop part  of (\ref{kloi}) then reads
\begin{multline}
\expval{J^i_\fl}{\Opr}{\lambda\Lambda}=-2\ii \eo^2\spinz
\int\frac{\dd{p}}{(2\pi)^4}\frac{2(p^0-\mo)^2 + (p_1)^2+(p_2)^2}{[(p-f)^2-\mphh+\izero]^2}\\
\cdot\B{\frac{1}{p^2-m_\o^2+\izero}-\frac{1}{p^2-\Lambda^2+\izero}}.
\label{kloi_PV}
\end{multline}

To evaluate it, we join the propagators' denominators  through the formula
\be
\frac{1}{AB^2}=\int_0^1 da\, db\, \delta(a+b-1)\frac{2b}{(a A + b B)^3},
\ee
shift the integration variable to make the resulting denominator 
$p^2$-dependent, Lorentz-average  the numerator of the integrand with
\be
p^\mu p^\nu\to \frac{\eta^{\mu\nu}}{4}p^2,
\ee
and perform Wick rotation  to arrive at   
\begin{subequations}
\begin{align}
&\expval{J^i_\fl}{\Opr}{\lambda\Lambda}=-\frac{\eo^2\spinz}{8\pi^2}
[I(\lambda,\mo)-I(\lambda,\Lambda)],\\
&I(\lambda,\chi)=\int_0^1 ds\frac{2s(1-s)^2}{(1-s)[(\chi/\mo)^2-s]+s(\lambda/\mo)^2}.
\end{align}
\label{hhnnn}%
\end{subequations}
Combining this  with
\be
\lim_{\lambda\to0}I(\lambda,\mo)=1, \ \limLmph I(\lambda,\Lambda)=0,
\ee
which can be straightforwardly shown, we finally get
\be
\exval{J^i_\fl}{\Opr}=
\limLmph\expval{J^i_\fl}{\Opr}{\lambda\Lambda}+O(\eo^4)=
-\spinz\frac{\eo^2}{8\pi^2}+O(\eo^4).
\label{JPVbar}
\ee
The same result is obtained if one first integrates (\ref{kloi_PV}) 
over $p^0$ using the residue theorem and then performs radial and angular
integrations in the $\p$-space in an arbitrary order.

\section{Discussion}
\label{Discussion_sec}
We have shown that a finite value for angular momentum stored in electric and magnetic   fields
of the electron can
be obtained  in quantum electrodynamics.
This is a non-trivial result because   individual 
components of  electron's angular momentum  need not be finite 
\cite{BurkardtPRD2009,LiuPRD2015,JiPRD2016}. 
Interestingly, our 
calculations of this  fundamentally-important not-so-intuitive 
quantity have not employed  any renormalization
procedure.

The complication, which we have encountered, is that we have actually obtained two
finite one-loop results for   field angular momentum of the electron:
(\ref{qqqmmm}) and (\ref{JPVbar}) in 3D cutoff- and Pauli-Villars-regularized
QED. 
Using $\eo=e+O(e^3)$, they   can be written as
\begin{subequations}
\be
\exval{J^i_\fl}{\Opr}=-\spinz\frac{\alpha}{3\pi}+O(\alpha^2)
\label{Jcutfin}
\ee
and
\be
\exval{J^i_\fl}{\Opr}=-\spinz\frac{\alpha}{2\pi}+O(\alpha^2),
\label{JPVfin}
\ee
\label{twoJ}%
\end{subequations}
respectively.

We suspect that the disagreement is caused by the lack of recovery of the
Lorentz symmetry upon removal of the 3D cutoff regularization. Such a
regularization, unlike the Pauli-Villars regularization, 
explicitely breaks this  symmetry in the intermediate steps of the
calculations. 
As a result, we are inclined to think that Pauli-Villars-regularized result (\ref{JPVfin}) provides 
the correct value of   field angular momentum of the electron. 
At the same time, we  hope that these two findings  will
 stimulate   discussion of  regularization (in)dependence of QED
calculations. We also hope that they will
motivate experimental studies of field angular momentum of
the electron.  

These  results can be now compared to  the
classical estimation that we have discussed in Sec. \ref{Introduction_sec}.
Such a comparision is of  interest if one aims at  getting intuitive
insights into  the QED calculations.
We find two curious  differences between  (\ref{Jclass2}) and (\ref{twoJ}). 

First, (\ref{Jclass2}) overestimates field 
angular momentum of the electron by  roughly  three orders of magnitude.

Second,  field angular momentum of the electron  is
anti-aligned with the electron's  spin in (\ref{twoJ}). The opposite is observed 
in (\ref{Jclass2}). This is illustrated  in  Fig. \ref{zeb_fig},
where the assumed downward orientation of the magnetic moment $\bmu$ implies upward
orientation of the spin of a negatively charged particle.

The first difference can be made less severe  by increasing the cutoff $r_c$. For example, 
one may try 
\be
r_c=O(r_0) \to O\B{\frac{r_0}{\alpha}}.
\ee
This modification makes  sense because  QED corrections 
to the Coulomb field are non-negligible at distances smaller  than the
reduced Compton wavelength, which is given by $1/m=r_0/\alpha$ \cite{PS}.
In other words, the classically-motivated cutoff used in Sec.
\ref{Introduction_sec} leads to employment of 
 expression (\ref{EBclass}) for the Coulomb   field   well beyond its  range of applicability.

Such a fix, however, has no influence on the second difference.
If we now assume  that   classical expression (\ref{Jclass2}) 
 captures long-distance contribution to field angular momentum of the electron, we
could conclude from (\ref{twoJ}) that the short-distance 
contribution to this quantity   is crucial for getting the right answer. 
This is the reason why classical estimations, akin to what we have presented 
in Sec. \ref{Introduction_sec}, will always have to be incomplete.

Next, at the risk of stating the obvious, we mention   that it would be 
most desirable to have an experimental measurement of   field angular
momentum of the electron. Given the fact that we deal here with a  gauge invariant 
observable, whose expectation value is finite,
it seems reasonable to assume that  such a measurement may  be feasible. 
Perhaps one difficulty associated with it  would be that the quantity of
interest here   is rather small. The same, however,
can be said about the Schwinger's correction to the 
electron's magnetic moment, which was measured about seven decades ago (see 
e.g. \cite{ComminsAnnRev2012}).
Therefore, the big open question is how one can 
experimentally approach such a quantity.

\noindent{\bf Acknowledgements}\\ 
\noindent 
I would like to thank Aneta for being a wonderful sounding board during
all these studies. I would also like to thank the Referee for his/her remarks 
about  the lack of the $\sz$-independent component  of field angular momentum of the electron.
This work is supported by the Polish National Science Centre (NCN) grant DEC-2016/23/B/ST3/01152.

\appendix

\section{Conventions}
\label{Conventions_sec}
We use the Minkowski metric $\eta=\text{diag}(+---)$ and choose $\varepsilon^{0123}=+1=\varepsilon^{123}$.
Greek and Latin indices  take values $0,1,2,3$ and   $1,2,3$, respectively,
when they refer to the components of $4$- and $3$-vectors. 
The  Einstein summation convention is applied to those indices. Moreover, 
$3$-vectors are written  in bold,  e.g. $x=(x^\mu)=(x^0,\x)$.

We employ Heaviside-Lorentz  units and set  $\hbar=c=1$. The fine-structure
constant is then given by 
\be
\alpha=\frac{e^2}{4\pi}.
\ee

We work in the standard representation of $\gamma$ matrices.
The normalization condition of single-electron eigenstates of the Dirac Hamiltonian, 
say $|\p s\ra$ with $\p$ being the electron's   $3$-momentum and $s$ being its
spin state, is $\la\p s|\p's'\ra=\delta(\p-\p')\delta_{ss'}$.
The $\us$ bispinors, which  appear in contractions on external lines, are normalized such that  
\be
  \ \us=\left(
\begin{array}{l}
1\\0\\0\\0
\end{array}
\right) \ \text{for} \ \sz=+1/2,  \ 
\us=\left(
\begin{array}{l}
0\\1\\0\\0
\end{array}
\right) \ \text{for} \ \sz=-1/2. 
\label{u}
\ee
They are eigenstates of the $z$-component of the one-particle fermionic spin angular momentum
operator,  $\ii\varepsilon^{3mn}\gamma^m\gamma^n/4$,
to the eigenvalue $\sz$. Moreover, $(\ii\gamma^\mu\partial_\mu-\mo)e^{-\ii\mo t}\us=0$.

\end{document}

%% file: sheader
\def\x{\boldsymbol{x}}
\def\y{\boldsymbol{y}}
\def\z{{\boldsymbol z}}
\def\o{{\text{o}}}

\def\0{\boldsymbol{0}}

\def\J{\boldsymbol{J}}
\def\dJ{\mathfrak{J}}

\def\JJ{{\boldsymbol\dJ}}

\def\q{\boldsymbol{q}}
\def\p{{\boldsymbol{p}}}

\def\r{\boldsymbol{r}}

\def\E{\boldsymbol{E}}

\def\Bold{\boldsymbol{B}}
\def\P{\boldsymbol{P}}

\def\T{\mathds{T}}

\def\ii{\mathrm{i}}

\def\d#1{d^3\mkern-1.5mu#1\,}

\def\dd#1{d^4\mkern-1.5mu#1\,}
\def\DD#1{[d^4\mkern-1.5mu#1]}

\def\ddd#1{\frac{d^3\mkern-1.5mu#1}{(2\pi)^3}}
\def\dddd#1{\frac{d^4\mkern-1.5mu#1}{(2\pi)^4}}

\def\bmu{{\boldsymbol\mu}}

\newcommand{\rfock}[3]{
\ifthenelse{\equal{#2}{}}{|#1,\lfloor#3\rfloor\ra}{|#1,\lceil#2\rceil,\lfloor#3\rfloor\ra}
}
\newcommand{\lfock}[3]{
\ifthenelse{\equal{#2}{}}{\la#1,\lfloor#3\rfloor|}{\la#1,\lceil#2\rceil,\lfloor#3\rfloor|}
}

\def\varimn{\varepsilon^{imn}}

\def\eo{e_\o}
\def\mo{m_\o}

\def\vareps#1{\varepsilon_{\boldsymbol{#1}}}

\def\om#1{\omega_{\boldsymbol{#1}}}

\def\xxii{{\frac{1-\xi}{\xi}}}

\def\asym{\text{asym}}
\def\sym{\text{sym}}

\def\IN{{\rm int}}

\def\la{\langle}
\def\ra{\rangle}

\def\limT{\lim_{T\to\infty(1-\ii0)}} 
\def\limlambdyc{\lim_{\substack{\Lambda_c\to\infty \\ \lambda_c\to0}}}
\def\intTT{\int^T_{-T}}
\def\intT{\int_T}

\def\barexymunu{x,\mu\leftrightarrow y,\nu}

\def\us{u_s}
\def\ous{\overline{u}_s}

\def\B#1{\!\left(#1\right)}
\def\BB#1{\!\left[#1\right]}

\def\N#1{:\!#1\!:}

\def\exval#1#2{\la#1\ra_{#2}}
\def\expval#1#2#3{\la#1\ra_{#2}^{#3}}

\def\trr#1{\text{Tr}\!\left[#1\right]}

\def\be{\begin{equation}}
\def\ee{\end{equation}}

\def\bee{\begin{equation*}}
\def\eee{\end{equation*}}

\def\bg{\begin{equation}\begin{gathered}}
\def\eg{\end{gathered}\end{equation}}

\def\bgg{\begin{equation*}\begin{gathered}}
\def\egg{\end{gathered}\end{equation*}}

\def\ba{\begin{equation}\begin{aligned}}
\def\ea{\end{aligned}\end{equation}}

\def\mph{\lambda}
\def\mphh{\lambda^2}

\def\limLmph{\lim_{\substack{\Lambda\to\infty \\ \mph\to0}}}

\def\izero{\ii0}

\def\sz{s_z}
\def\spinz{\sz\delta^{i3}}

\def\fl{\text{field}}

\def\Opr{\boldsymbol{\Omega}s}
\def\Oprim{|\Opr\ra}

\def\Vol{V}

\makeatletter
\newcommand{\pushright}[1]{\ifmeasuring@#1\else\omit\hfill$\displaystyle#1$\fi\ignorespaces}
\newcommand{\pushleft}[1]{\ifmeasuring@#1\else\omit$\displaystyle#1$\hfill\fi\ignorespaces}
\makeatother

\def\XXint#1#2#3{{\setbox0=\hbox{$#1{#2#3}{\int}$}
\vcenter{\hbox{$#2#3$}}\kern-.5\wd0}}

\newcommand{\Rzymskie}[1]{%
  \textup{\uppercase\expandafter{\romannumeral#1}}%
}